\documentstyle[12pt,epsf,epsfig,a4wide]{article}

\textwidth17truecm \textheight22truecm \fontsize{12}{14pt} \selectfont

\begin{document}

\def\beq{\begin{equation}}
\def\eeq{\end{equation}}
\newcommand{\form}[1]{(\ref{#1})}

\begin{centering}

\begin{flushright}
astro-ph/0401532 \\
CERN-PH-TH/2004-004 \\
January 2004
\end{flushright}

\vspace{0.1in}

{\Large\bf Primordial Structure of Massive Black Hole Clusters}

\vspace{0.4in}

{{\bf M.Yu.~Khlopov$^{a,b,c}$}, {\bf S.G.~Rubin$^{a,b}$} and {\bf 
A.S.~Sakharov$^{d,e,f}$}}
\\
\vspace{0.2in}

\begin{flushleft}

$^a$ Centre for CosmoParticle Physics "Cosmion",  4 Miusskaya pl., 125047
Moscow, Russia
\\
$^b$ Moscow Engineering Physics Institute, Kashirskoe shosse 31, 115409
Moscow, Russia \\
$^c$ Physics Department, University Rome~1 La Sapienza, Ple A. Moro 
2, 
00185 Rome, Italy \\
$^d$ Department of Physics, CERN
Theory Division, CH--1211 Geneva 23, Switzerland \\
$^e$ Swiss Institute of Technology, ETH-Z\"urich, CH--8093 Z\"urich, 
Switzerland
\\
$^f$ INFN Laboratory Nazionali del Gran Sasso, SS. 17bis 67010 Assergi 
(L'Aquila), Italy

\end{flushleft}

\vspace{0.4in}
 {\bf Abstract}

\end{centering}

\vspace{0.5in}

{\small \noindent We describe a mechanism of the primordial black holes
formation that can explain the existence of a population of supermassive black
holes in galactic bulges. The mechanism is based on the formation of black
holes from closed domain walls. The origin of such domain walls could be a
result of the evolution of an effectively massless scalar field during 
inflation.
The initial non-equilibrium distribution of the scalar field imposed by
background de-Sitter fluctuations gives rise to the spectrum of black holes,
which covers a wide range of masses -- from superheavy ones down to deeply
subsolar. The primordial black holes of smaller masses are concentrated around
the most massive ones within a fractal-like cluster.}

\vspace{0.8in}
\begin{flushleft}
CERN-PH-TH/2004-004 \\
January 2004
\end{flushleft}



\newpage

\section{\label{introduction}Introduction}

There is growing evidence for the existence of massive dark objects at the
centers of galaxies, which are presumably supermassive black holes (SMBHs) of
mass $10^6$--$10^{10}M_{\odot}$. The existence of SMBHs was predicted by
arguments based on the active galactic nuclei (AGNs) and quasistellar objects
(QSOs) energetics and demography~\cite{zeldbh,rees1,kormgeb}. Also a variety of
techniques exists in the mean time to detect and weigh SMBHs, using
measurements of stellar or gaseous kinematics~\cite{kormgeb}. This has
revealed that many, and may be all, quiescent galaxies host SMBHs as well~%
\cite{rich}, including the Milky Way~\cite{mw}. The mass of a SMBH is always
of the order of 0.1\% of the galaxy bulge/spheroid mass~\cite{kormgeb}. An
even better correlation exists between the mass of SMBHs and the stellar
velocity dispersion in the host galaxy, $M_{BH}\propto\sigma^4$~\cite%
{gebhart}, which holds 3--4 orders of magnitude.

The origin of SMBHs is far from certain, but several theories have been
advanced. The above-mentioned close relationship between BH masses and the
stellar velocities dispersion may be interpreted in various ways, and
ultimately one is confronted with the problem of what came first, the host
galaxy or the SMBHs. It is not inconceivable that SMBHs are purely the result
of the galactic dynamics and their merging history; yet no one has proposed a
concrete mechanism for converting stellar mass objects into others 6--10 orders
of magnitude larger. In particular the paradigm of the formation of SMBHs as a
pure result of galactic dynamics is hard to match with the discovery of quasars
at redshift $z\ge 6$~\cite{highzq}. The point is that the formation of SMBHs
through the growth of BH seeds by gas accretion is supported by the consistency
between the total density in quasars light and the SMBH mass density in local
galaxies, adopting a reasonable accretion rest-mass-to-energy conversion
efficiency~\cite{acreff}. From the other side, SMBHs must be present by $z\ge
6$ to power the above-mentioned QSOs. It has been argued~\cite{gnedin1} that if
they grew by accretion from smaller seeds, substantial seeds of mass $\ge
10^5M_{\odot}$ must have been present already at $z\approx 9$ to have
sufficient time to grow up to a typical quasar black hole mass $\simeq
10^9M_{\odot}$. Thus a very massive progenitor at $z\approx 9$ is very likely.
At that epoch large galaxies do not yet exist. In the hierarchical formation
scenario, they assemble via infall and merging of small clumps of baryonic gas
and dark matter. Under such circumstances it seems that in order to form a huge
BH in a gasodynamical process, two generic conditions must be satisfied, namely
the gas must be strongly self-gravitating; and star formation must be
inefficient, allowing a large supply of cold gas to be converted to a BH. The
second constraint is of particular importance, since star formation is a
competitor of black hole formation. So it is clear that to prevent the
formation of stars, small-scale density perturbations should be effectively
washed out or damped. A way to achieve this is not trivial. Indeed the power
CDM spectrum for lightest supersymmetric particles, for instance, at
subgalactic scales~\cite{schwarz} implies that at $z>9$ all fluctuations below
the mass scale $10^8M_{\odot}$ went to the non-linear regime, requiring some
special mechanisms to dilute small scale density
perturbations and consequently suppress star formation processes~\cite%
{gnedin1,shapiro}. So, summing up the above arguments it seems appealing
that central primordial objects of mass $\ge10^5M_{\odot}$ preceded the
formation of the first galactic bulges and the beginning of quasar activity.

Previously~\cite{we1,jetp,inhom} we proposed a mechanism for the formation of
the massive primordial BHs (PBHs), which opens a possibility that primordial
objects of the mass $\ge 10^2M_{\odot}$ were created before any galaxy
formation processes started to develop. The mechanism is based on the fact that
BHs can be created as a result of a collapse of closed vacuum walls formed
during a second-order phase transition with strongly non-equilibrium initial
conditions. The required initial conditions get generated in inflationary
dynamics of a scalar field, with a potential possessing an effectively flat
valley~\cite{we1}. A wide class of models possessing that symmetry-breaking
pattern can be effectively described by a pseudo-Nambu--Goldstone (PNG) field
and corresponds to the formation of an unstable topological defect in the early
Universe. In particular, PNG fields received substantial attention in
connection with curvaton proposal~\cite{curv}.

The mechanism~\cite{we1,jetp,inhom} assumes the succession of phase
transitions, related with the consequence of spontaneous and manifest U(1)
symmetry breaking. The crucial point is that in the considered case spontaneous
symmetry breaking takes place during inflation so that neither angular nor
radial fluctuations can lead to formation of strings. It makes closed walls the
only topological defects, which are formed after manifest breaking of the
underlying U(1) symmetry. For a wide range of realistic parameters of the
considered model closed walls turn out to be stable relative to the formation
of holes bounded by strings in them. The distribution of walls is also
practically not changed by wall collisions. It makes possible to deduce the
spectrum and space distribution of BHs from the distribution of closed walls
formed after the manifest U(1) symmetry breaking.

Actually it has long been known \cite%
{kp,Khlopov85,Novikov79,3,4,mi,ck,klopmop} that the PBH formation is
possible in the early Universe. PBH can form if the cosmological expansion
of a mass stops within its gravitational radius, which happens when the
density fluctuation exceeds unity on a scale between the Jeans length and
the horizon size. Other possibilities are related to the dynamics of various
topological defects such as the collapse of cosmic strings \cite{3} from the
thermal second--order phase transition or to the collisions of the bubble
walls \cite{4,mi} created at the first--order phase transitions. Formally,
there is no limit on the mass of a PBH that forms when a highly overdense
region collapses: it is only needed to form, at inflation, the spectrum of
initial density fluctuations that provides the proper inhomogeneity at the
desired scale~\cite{lk}. From the other side, to preserve the successful
predictions of the Big Bang nucleosynthesis (BBN) and general features of
the standard Big Bang scenario, confirmed by observations (see~\cite{klopmop}
for a review), there is no room for significant cosmological inhomogeneity
once nucleosynthesis begins; the PBHs therefore seem not to have
sufficiently large masses, being more likely to represent the specific form
of a modern collisionless dark matter, rather than collecting into a large
BH, which could play the role of a seed for the galactic nuclei.

The mechanism described in~\cite{we1} offers a way around this, allowing the
formation of closed vacuum walls of a size significantly exceeding the
cosmological horizon at the moment of their formation. At some instant after
crossing the horizon, such walls become causally connected as a whole and
begin to contract, owing to their surface tension. This contraction ends up
with the formation of a BH with a mass of the same order as that of the
wall, when it enters the horizon. Therefore the size distribution of closed
vacuum walls gets converted into the mass spectrum of resulting PBHs. Such a
spectrum causes no problem for BBN predictions and satisfies all the other
restrictions of possible PBHs contributions~\cite{we1}, while providing a
sufficient number of very massive BHs that could seed galactic nuclei%
\footnote{%
Another possibility of massive PBH formation is considered in~\cite{quin};
it involves the blue initial spectrum of fluctuation and subsequent
accretion of quintessence. However, the last measurements of cosmological
parameters by WMAP~\cite{wmap} disfavor the blue spectrum of initial
perturbations. These measurements do not exclude the possibility of massive
PBH formation by spikes on the background of the red spectrum, which can be
realized in the mechanism~\cite{lk}.}. The other generic feature of the
mechanism~\cite{we1} is the fractal structure of initial domain walls, which
reproduces itself in a more and more decreasing scale. After the domain
walls collapse, their fractal distribution, generated in fact during
inflation gets imprinted into space distribution of less massive PBHs around
the most massive ones.

In the present paper we develop the ideas firstly revealed in~\cite%
{we1,jetp} and investigate the space correlation properties of the distribution
of PBHs, which can modify the power spectrum and can lead to the fractal-like
correlation properties observed in the galaxy distribution at least up to
scales 100~Mpc~\cite{corrfun}. The paper is organized in the following way. In
the second section we describe the main issues of the mechanism~\cite{we1} and
discuss observational data in favor of early BH formation. The third section
deals with the spatial distribution of the resulting clouds of PBHs. Finally,
in section 4 we conclude and discuss possible effects of fractal-like PBH
distribution.

\section{\label{formation} Formation of PBHs in inflationary PNG dynamics}

We review here a mechanism of primordial SMBHs~\cite{we1} formation. This
mechanism 
is based on the inflationary dynamics of a
scalar field with two different vacuum states which get unequally 
populated after the end
of inflation deeply in the Friedman-Robertson-Walker (FRW) epoch. Such a
situation can occur quite naturally under the condition that the  potential of
the scalar fields possess a valley along which the effective  mass of the field
vanishes comparable to the Hubble rate during inflation. The latter causes a
large friction term  in the equation of motion of the scalar field making it in
fact to be effectively massless during inflation. The subsequent exit from the
inflationary regime and further cooling of the Universe lead to decreasing of
the expansion rate and therefore making the scalar field moving freely toward
the minima of its potential. However it is the ability of quantum fluctuations
of effectively massless scalar field on the de-Sitter
background~\cite{Star79} to redistribute the scalar field unequally between two
different states. It makes the Universe
to contain islands of  less probable vacuum, surrounded by the sea of another,
more preferable one. This means that once the vacuum states are populated at
some time in FRW epoch, closed vacuum walls should appear in the Universe, 
surrounding the  islands of less probable vacuum.
When a closed domain wall becomes causally connected, it starts to collapse 
forming
a BH with mass, being equal to that of the progeniting vacuum wall. It
was shown in~\cite{we1} that this mechanism can provide formation of
sufficiently large closed vacuum walls, so that primordial SMBHs cold be formed 
with masses
and abundance satisfying observations.

Below we use the working example of a complex
scalar field with the potential
\begin{equation}  \label{V1}
V(\varphi ) = \lambda (\left| \varphi \right|^2 - f^2 /2)^2+\delta V(\theta
),
\end{equation}
considered in~\cite{we1}, where $\varphi = r_{\varphi}e^{i\theta } $. This
field is taken to be coexisting with the inflaton that drives the Hubble
constant $H$ during the inflationary stage. In this paper we accept the value
$H=10^{13}$~GeV for the Hubble constant during inflation, which is consistent
with resent results of WMAP~\cite{infl5} on the energy scale of inflation. The
term
\begin{equation}  \label{L1}
\delta V(\theta ) = \Lambda ^4 \left( {1 - \cos \theta } \right),
\end{equation}
which reflects the contribution of instanton effects to the Lagrangian
renormalization (see for example \cite{adams}), is negligible during the
inflationary stage and during some period in the FRW expansion. In other words,
the parameter $\Lambda$ vanishes with respect to $H$. The omitted term
(\ref{L1}) induces the effective mass of an angular field, and it begins to
play a significant role only at the moment, after inflation, when the Hubble
parameter is equal to this effective mass in its gradual decrease with time ($H
= 1/2t$ during the radiation-dominated epoch). Also, we assumed in~\cite{we1}
the mass $m_r =\sqrt{\lambda}f$ of the radial field component $r_{\varphi}$ to
be always sufficiently large with respect to $H$, which means that the complex
field is in the ground state even before the end of inflation. Since the term
(\ref{L1}) is negligible during inflation, the field has the form $\varphi
\approx f/\sqrt 2 \cdot e^{i\theta } $, the quantity $f\theta$ acquiring the
meaning of a massless angular field.

In~\cite{we1} we applied the fact that the quantum fluctuations of the phase
Sitter background can be pictured as a one-dimensional
Brownian motion along the circular valley of the potential (\ref{V1}): the
magnitude of $\theta$ on the scale of the de Sitter horizon ($H^{-1}$)
fluctuates by an amount
\begin{equation}  \label{fluctphase}
\delta \theta = H/2\pi f
\end{equation}
per Hubble time ($H^{-1}$). The amplitude of those fluctuations with
wavelengths exceeding $H^{-1}$ freezes out at its value because of the large
friction term in the equation of motion of the phase, whereas its wavelength
grows exponentially. Such a frozen fluctuation is equivalent to the appearance
after every $e$-fold of a classical increment/decrement of the phase by factor
$\delta\theta$. Therefore assuming that the inflation begins in a single
causally connected domain of size $H^{-1}$, which contains some average value
of phase  $\theta_0<\pi$ one can conclude that more and more domains appear
with time, in which the phase differs significantly from the initial value
$\theta_0$. A principally important point of~\cite{we1} is the appearance of
domains with phase $\theta >\pi$. Appearing only after a certain period of time
($e$-foldings) during which the Universe exhibited exponential expansion, these
domains turn out to be surrounded by a space with phase $\theta <\pi$.

\begin{figure}[t]
\begin{center}
\epsfig{file=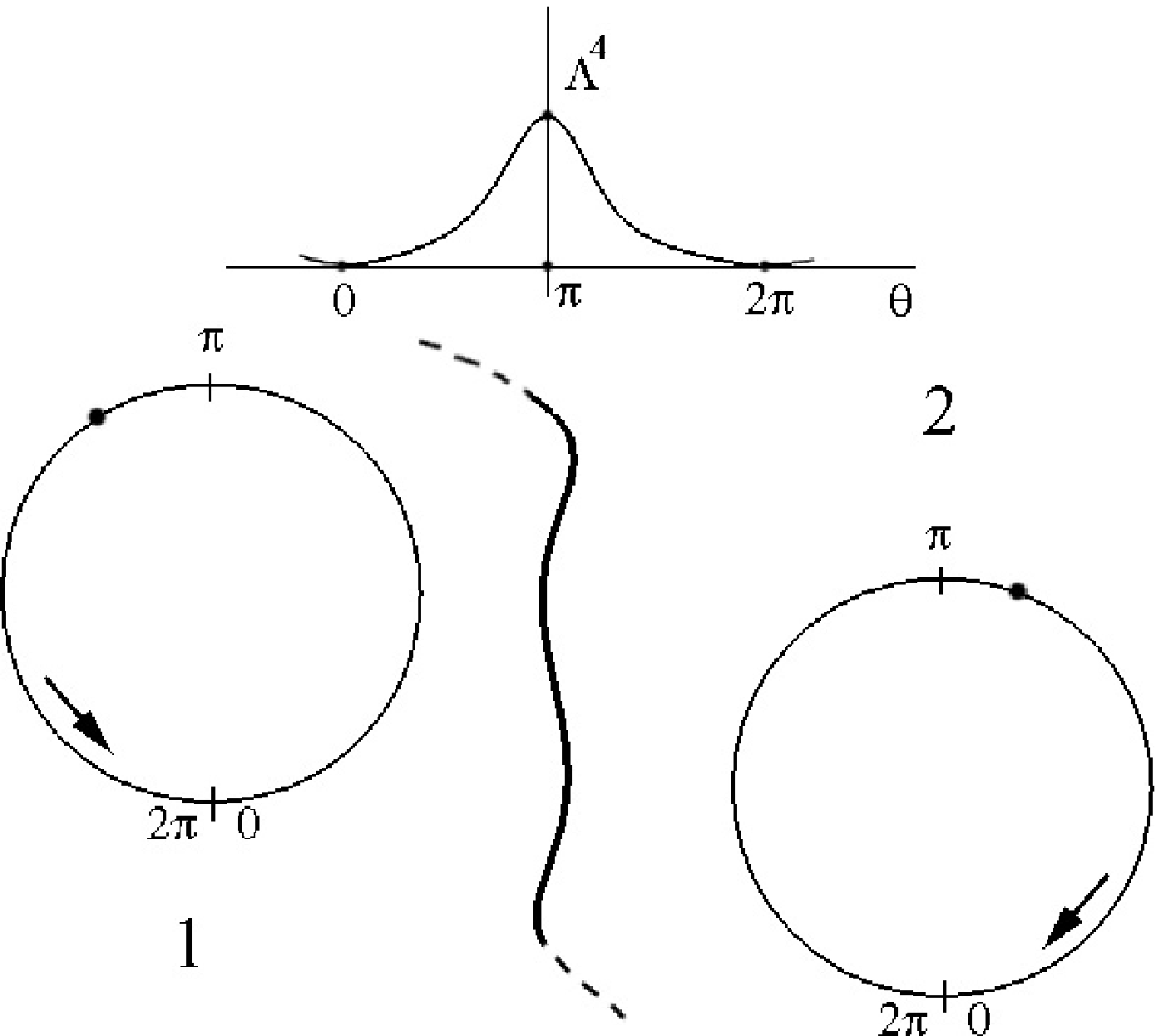,width=0.8\textwidth}
\end{center}
\caption{\textit{The explicit deconvolution of the potential (\protect\ref%
{L1}) (top) shows that the vacuum manifold of angular component $\protect%
\theta$ consists of two stable states associated with $0$ and $2\protect\pi$
akin to the simple double-well potential (only the two closest minima are
shown). This situation is an analogue of that with the breaking of discrete
symmetry. A domain wall occurs at the boundaries, between regions of space with
the vacuum value of the phase $\protect\theta$ in different states, with
$\protect\theta$ interpolating between $0$ and $2\protect\pi$ across the wall.
The energy of the configuration is rising over the potential hill. Two
neighboring domains (bottom) filled with the phase values above and below
$\protect\pi$, corresponding to the two different vacuum states, give rise to
the formation of a domain wall once the motion of the angular component becomes
underdumped, driving the phase to different minima, namely $2\protect\pi$ and
$0$ respectively.}} \label{wall}
\end{figure}

The potential (\ref{V1}) possesses a $U(1)$ symmetry, which is spontaneously
broken, at least after some period of inflation. This spontaneous breakdown
is provided by the condition of sufficiently large radial mass, $m_r=\sqrt{%
\lambda_{\phi}}f>H$. At the same time the condition
\begin{equation}  \label{angularmass}
m_{\theta}=\frac{f}{\Lambda}^2\ll H
\end{equation}
on the angular mass provides the freezing out of the phase distribution
until some moment of the FRW epoch. After the violation of condition (\ref%
{angularmass}), the term (\ref{L1}) contributes significantly to the
potential (\ref{V1}) and explicitly breaks the continuous symmetry along the
angular direction. Thus, potential (\ref{V1}) eventually has a number of
discrete degenerate minima in the angular direction at the points $%
\theta_{min}=0,\ \pm 2\pi ,\ \pm 4\pi,\ ...$ . As soon as the angular mass
$m_{\theta}$ is of the order of the Hubble rate, the phase starts be
oscillating~\footnote{The discussion on the coherent phase oscillations in the
cosmology of the invisible axion see in \cite{kim} and references therein.
Cosmology of rather heavy axion predicted in string theory is discussed in
\cite{Bozza:2002ad}. Variability of $f$, induced by inflaton, is discussed in
\cite{progress}.} about the potential minimum, initial values being different
in various space domains. Moreover, in the domains where the initial phase is
$\pi <\theta < 2\pi $, the oscillations proceed around the potential minimum at
$\theta_{min}=2\pi$, whereas the phase in the surrounding space tends to a
minimum at the point $\theta _{min}=0$. This implies that closed vacuum
walls~Fig.\ref{wall} get formed~\footnote{The existence of such domain walls in
the theory of invisible axion was first pointed out
in~\cite{sikivieinvisible}.} separating the island with $\theta_{min}=2\pi$
from those with $\theta _{min}=0$. The surface energy density of a wall of width $\sim
1/m\sim f/\Lambda^2$ is of the order of $\sim f\Lambda ^2$. Those walls, which
are larger than the cosmological horizon, still follow the general FRW
expansion until the moment when they get causally connected as a whole; this
happens as soon as the size of a wall
becomes equal to the horizon size $R_h$ \footnote{%
This condition can be violated for very large domains, as discussed in the next
section.}. Evidently, internal stresses developed in the wall after crossing
the horizon initiate processes tending to minimize the wall surface. This
implies that the wall tends, first, to acquire a spherical
shape and, second, to contract toward the centre~\footnote{%
The motion of closed vacuum walls has been derived analytically in \cite%
{tkachev,sikivie}.}. Should the wall at the same moment be localized within the
gravitational radius, a PBH is formed. The phase distribution generated during
inflation is reflected in the size distribution of vacuum walls and hence in
the mass distribution of BHs appearing after the walls collapse.

The above considerations are valid under the assumption that $r_{\varphi}$
is situated at the vacuum value $r_{\varphi}=f$ during the last 60
e-foldings of inflation, so that the only degree of freedom is the massless
angular component $f\theta$. One can take into account the radial
fluctuations which can actually kick $r_{\varphi}$ out of the vacuum. The
consequence of $r_{\varphi}$ being kicked out of the vacuum over the top of
the Mexican hat potential $V(\varphi )$ is that strings get formed along the
lines in space where $r_{\varphi}=0$. If it is the case the usual thermal
scenario of the formation of the system of topological defects walls bounded
by strings akin we have in the theory of invisible axion gets realized.
Subsequently such structure of walls bounded by strings would disappear
rapidly due to disintegration into separate fragments and futher emission of
axion-like particles (see for review~\cite{kim}). However the situation with
radial fluctuations has been investigated both numerically and analytically,
using the relevant Langeven and Fokker-Planck equations~\cite{luth}. In
particular it was found that $r_{\varphi}$ is kicked out of the vacuum if
the Hubble constant during inflation exceeds the radius of the bottom of the
Mexican hat potential $H\ge f$. In that case, strings form and the thermal
scenario is reproduced. In some sense there is a ``Hawking temperature'' of
order $H$ during inflation which takes over to drive thermal-like phase
transition. From the other side, as it was found in~\cite{luth}, $%
r_{\varphi} $ sits in the vacuum whenever $H\le f$ during inflation. The
last case is valid for the reasonable value of $\lambda\simeq 10^{-2}$ and
simplest exponential inflation. Therefore in the regime $f\approx 10H$
considered in present paper (see numerical parameters in Section \ref%
{distribution}) the only degree of freedom is the angular component $f\theta$%
, so only closed vacuum walls do form while the formation of strings is
prevented.

Generally speaking the domain walls we discuss, can decay by quantum nucleation
of holes bounded by strings~\cite{nucleation,vs}. Semiclassically, the hole
nucleation is described by an instanton and the decay probability can be
expressed as~\cite{nucleation,vs}
\begin{equation}  \label{nuclprob}
P=A\exp \left(-\frac{16\pi\mu^3}{3\sigma^2}\right),
\end{equation}
where $\mu\simeq\pi f^2$ and $\sigma\simeq f\Lambda^2$ are the string and
wall tensions, respectively, while the pre-exponent factor $A$ can be
calculated by analyzing small perturbations about the instanton~\cite%
{coleman}. Experessing (\ref{nuclprob}) via parameters of (\ref{V1}) and (%
\ref{L1}) one can easily see that the probability to create a hole is
suppressed by factor $\exp (-10^2(f/\Lambda)^4)$, which turns out to be very
small owing to large hierarchy $f/\Lambda\simeq 10^{11}$ we use to realize
the proposed mechanism (see the next section). Therefore on the timescale of
the age of the Universe the walls are effectively absolutely stable relative
to quantum nucleation of holes bounded by strings~\cite{nucleation,vs}.

There is also very unlikely that the initially closed walls can become open due
to their mutual intersections. This argument can be confirmed by numerical
analyzes~\cite{nagasawa}, which demonstrated in the case of parallel
sine-Gordon walls that such walls pass freely through one another. So it is
unlikely that collisions of the walls leads to formation of holes and that
closed walls can become open in such collisions. Due to non-perfect initial
form of walls their flattening takes place in course of wall evolution. Such
flattening changes the masses of the walls and, consequently, influences the
mass distribution of black holes, which we discuss below. However, walls
intersections in the course of flattening do not change the topology of walls,
they do not convert closed walls into open ones bounded by strings.

In the considered case the first (spontaneous symmetry breaking) phase
transition takes place on the inflationary stage and neither angular nor radial
fluctuations can provide formation of strings. It makes closed walls the only
topological defect solution that can appear after manifest symmetry breaking,
when the angular mass is switched on. Such walls are not bounded by strings by
construction, and neither quantum hole nucleation nor mutual intersection can
change the closed topology of these walls. Domain walls can become open when a
string loop that binds a wall inside it hits a larger wall~\cite{vs,sikhag}.
Then pieces of two walls annihilate and a vacuum hole bounded by string is
produced. However as we argued above all the walls are initially closed and
have no strings on their boundary, so during intersections they will pass each
other without annihilation~\cite{nagasawa} thus preserving their closed
character.

\section{ Space distribution of PBHs}

\label{distribution}

The general conditions for the formation of primordial black holes have been
presented in~\cite{jetp}. In that paper we also performed calculations for
formation of walls of smaller sizes inside the wall of larger size. The
conclusion is that their total mass is only twice a mass of main wall, which
determines the mass of the largest black hole. BHs are formed as the result of
the closed walls evolution. Hence, total mass of black holes disposed closely
to the largest BH increases weakly its effective mass.

The present section contains calculation of BH spectrum around the main BH. The
size of area which we consider is much greater (in orders of magnitude) than
the size of the wall progeniting the largest BH. It contains a lot of smaller
walls what strongly increase total mass of the BH cluster. So that is of great
interest to study the properties of distribution
of BHs with intermediate masses around the most massive ones over those scales
which can be roughly attributed to the galaxy clusters scale.

General strategy is as follows: we choose the scale of some region containing
BH with the given value of mass; then we consider what the value of phase
$\theta_r$ must have been when this region was in a single horizon. We do that
by deciding what value of $\theta$ would have (on average) led to a single
black hole of the given mass. Then we use that value of $\theta_r$ to find the
density of BHs of smaller sizes in the considered region.

Let $M_{0}$ be the mass of most massive BH and $M$ the integrated mass of the
coexisting BHs inside a sphere of radius $r$ centered at the most massive BH.
It is convenient in our mechanism to trace the space distribution of resulting
BHs by the distribution in size of their progenitive closed domain walls. In
this sense, there are three scales related to our problem, namely the size of
the closed wall $R_{0}$, which finally gets converted into the most massive BH;
a size $R$ of a closed wall that becomes a BH of intermediate mass inside the
above-mentioned sphere; and the radius $r$ of the sphere itself. Every
mentioned scale can be attributed to the number of $e$-foldings left until the
end of inflation. We assign these numbers of $e$-foldings as $N_{0}$, $N$ and
$N_{r}$ respectively. It is clear that a region of size $r$ formed at the
inflationary stage will follow the expansion of the Universe, depending upon
the current equation of state, and hence we can approximately admit
\begin{equation}
r=r(N_{r},t)=r_{U}(t)e^{N_{r}-N_{U}},  \label{rsphere}
\end{equation}%
where $r_{U}(t)$ is the size of the Universe at time $t$. Let us assume that
the angular component of our field has the phase $\theta _{r}$, once a
region of size $r$ leaves the horizon when the Universe still has $N_{r}$ $e$%
-foldings to inflate. So, at some later time, characterized by $N_{0}<N_{r}$
$e$-foldings before the end of inflation, one has already $e^{3(N_{r}-N_{0})}
$ causally disconnected regions inside the volume created at $e$-folding $%
N_{r}$. As we saw in the previous section, the phase $\theta $ fluctuates at
each $e$-folding like a one-dimensional Brownian motion. According to our
mechanism, the biggest domain wall is formed once the phase $\theta $ overcomes
$\pi $ for the first time, being initially (at the beginning of inflation)
somewhere in the range $[0;\pi ]$ (see Fig. \ref{wall}). It is this wall that
collapses into the most massive BH serving as the centre of BHs cluster. By
assumption, only one of the above-mentioned domains contains a phase larger
than $\pi $ at $e$-folding number $N_{0}$. So the probability of finding the
phase within the interval $[\pi ;\pi \pm \delta \theta ]$ in one of the
causally disconnected regions at e-folding number $N_{0}$ is equal to $%
1/e^{3(N_{r}-N_{0})}$.

On the other hand, the well known formula \cite{Vil82,Linde82,Star82}

\begin{equation}
W(N,\theta )=\frac{1}{\sqrt{2\pi \left( N_{r}-N\right) }}\exp \left[ -\frac{%
(\theta _{r}-\theta )^{2}}{2\delta \theta ^{2}\left( N_{r}-N\right) }\right]
\label{gauss}
\end{equation}%
represents the probability to find phase $\theta $ in the interval $[\theta
;\theta \pm \delta \theta ]$ at arbitrary e-folding number $N$ within causally
disconnected volumes with the size of order of $1/H$  provided they had unique
phase $\theta _{r}$ at e-folding number $N_{r}$. The value $\delta \theta $ is
given by formula (\ref{fluctphase}). Thus the number of causally independent
domains acquiring the phase in the interval $[\pi ;\pi \pm \delta \theta ]$
just during the  e-folding number $N$ is

\begin{equation}
K(N)=W(N,\pi )e^{3(N_{r}-N)}.  \label{KN}
\end{equation}

The value of the phase $\theta _{r}$ can be expressed by equating number of
domains containing the phase in the interval $[\pi ;\pi \pm \delta \theta ]$ at
e-folding number $N_{0}$ to unity (by definition)

\[
K(N_{0})=1,
\]%
what gives%
\begin{equation}
\theta _{r}=\pi -\delta \theta \sqrt{\left( N_{r}-N_{0}\right) \left[
6\left( N_{r}-N_{0}\right) -\ln (2\pi \left( N_{r}-N_{0}\right) )\right] }.
\label{tetar}
\end{equation}

Combining Eqs. (\ref{gauss}), (\ref{KN}), (\ref{tetar}) the probability of
formation of walls with smaller masses at $e$-folding number $N$ can be
obtained as follows:
\begin{equation}
K\left( N\right) =\frac{\left[ 2\pi \left( N_{r}-N_{0}\right) \right] ^{%
\frac{N_{r}-N_{0}}{2\left( N_{r}-N\right) }}}{\sqrt{2\pi \left(
N_{r}-N_{0}\right) }}\exp \left[ -\frac{3\left( N_{r}-N_{0}\right) ^{2}}{%
N_{r}-N}+3(N_{r}-N)\right]   \label{WN}
\end{equation}

\begin{figure*}[tbp]
\includegraphics{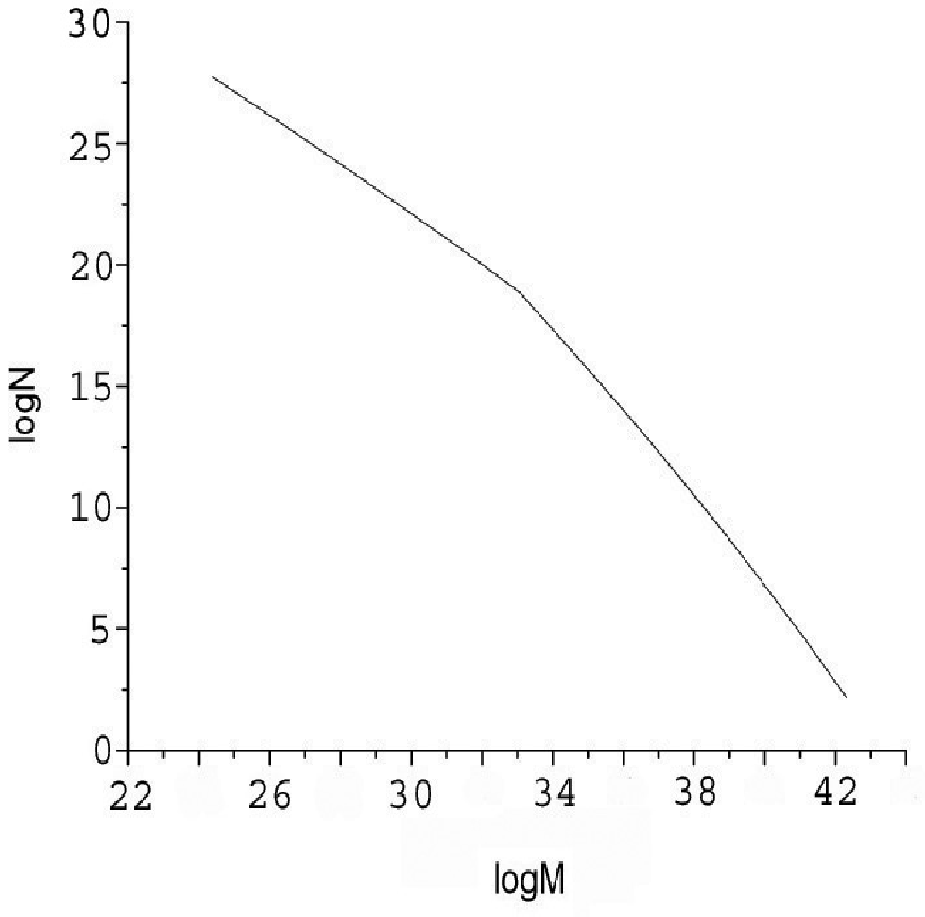}
\caption{\textit{Explicit mass (in g.) distribution of PBHs, simulated for the
following specific parameter: $\Lambda =100$ GeV, $f=8\times 10^{13}$ GeV. The
total mass of BHs over the Universe is $10^{53}$g.}} \label{massdist}
\end{figure*}

The formulae written above are formally valid for arbitrary parameter
$N_0$ which is connected with the mass of main BH. The BH of any mass
produces cluster of smaller BHs in the manner discussed above. The choice of
the mass of main BH could be done using mass distribution of BHs calculated
earlier - see Fig.\ref{massdist}. One can see from the figure, that amount of BHs
with mass $5\times 10^4{\rm M_{\odot}}$ - see Fig. \ref{massdist}- is about
$10^{11}$, i.e. approximately coincides with the total number of galaxies.
To treat the BH cluster structure as the pattern for future galaxy formation
we choose below the parameter $N_0$, corresponding to this value of mass for the
main BH. 

Now we shortly remind the connection between mass of black holes and e-folding
$N$ according to \cite{we1}. The mass $M$ of a closed wall is connected with
its size
\begin{equation}
M\approx 4\pi R\left( N\right) ^{2}\sigma ,  \label{wmass}
\end{equation}%
where $\sigma $ is the surface energy density of the wall.

\begin{figure*}[tbp]
\includegraphics{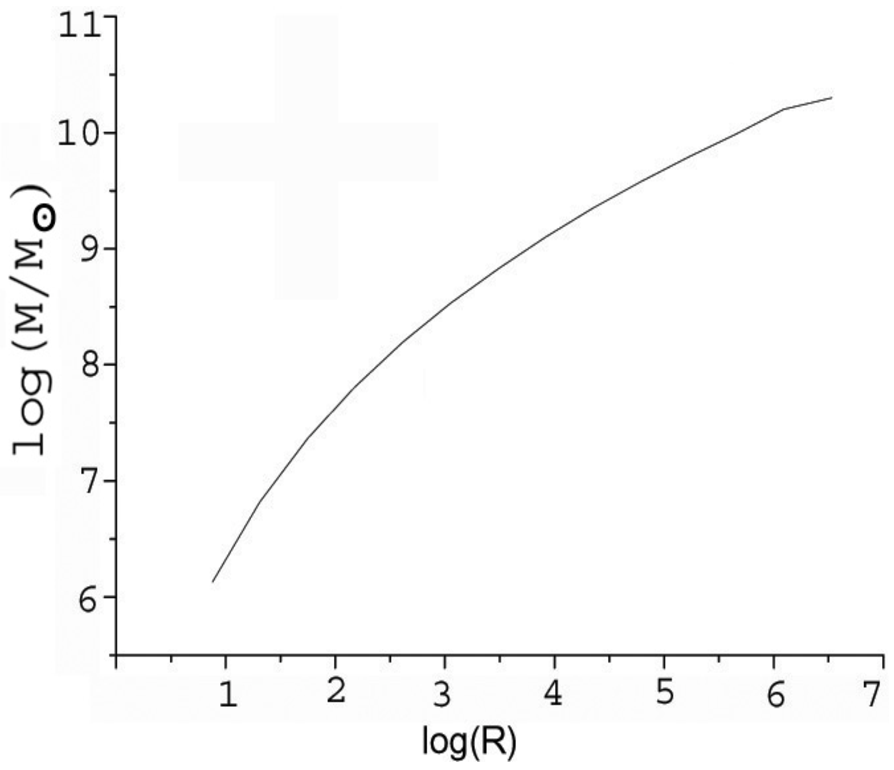}
\caption{\textit{Total mass of black holes within the sphere of radius $r$
(in pc.). Parameters of the model: $\Lambda =100$ GeV, $f=8\times 10^{13}$
GeV; the mass of the central BH is $2.7\times 10^4M_{\odot}$}}
\label{fig:radius}
\end{figure*}

There are two different scales related to the spherical wall dynamics. The
first one $ R_{1}(N)$ is determined from the condition that the wall mass
(\ref{wmass}) is equal to the energy of the plasma contained in the domain
bounded by this
closed wall at $t=R$, $E_{V}=\rho \frac{4\pi }{3}R^{3},$ where $\rho =\frac{%
\pi ^{2}}{30}g^{\ast }T^{4}$ is the energy density of the plasma with $%
g^{\ast }$ degrees of freedom during the radiation-dominated epoch. Taking
into account that
\begin{equation}
\sigma =4\Lambda ^{2}f,  \label{wallsurf}
\end{equation}%
and equating energies $E_{V}$ and $M$ from expression (\ref{wmass}), one
obtains the scale
\begin{equation}
R=\frac{3\sigma }{\rho },  \label{radcr}
\end{equation}%
which also can be expressed through the temperature $T$ when the
above-mentioned equality was established
\begin{equation}
R=L_{U}\frac{T_{0}}{T}e^{N-N_{U}};  \label{radtemp}
\end{equation}%
here $T_{0}$ and $L_{U}$ are temperature and size of the modern Universe.
Combining the formulas written above, one obtains the maximal size of
a spherical wall at which it does not influence the law of local
expansion, when it enters horizon:
\begin{equation}
R_{1}(N)\approx \left( \frac{\pi ^{2}}{90\sigma }g_{\ast }\right)
^{1/3}\left( L_{U}T_{0}\right) ^{4/3}e^{\frac{4}{3}(N-N_{U})}.
\label{limit1}
\end{equation}%
Recall that $L_{U}T_{0}\approx 2.7\times 10^{29}$.

Another scale arises if we take into account the internal pressure of the wall.
Its influence is developed after the wall comes under the horizon and leads to
the termination of the wall growing and its acquiring a spherical shape. This
moment is characterized by the instant $t_{2}$\ such that
\begin{equation}
l_{hor}(t_{2})=R(t_{2})\equiv R_2,  \label{lhor}
\end{equation}%
where the size of the horizon during the radiation-dominated era scales like
$l_{hor}(t_{2})=2t_{2}$. We also recall that walls get nucleated under the
violation of condition (\ref{angularmass}), which happens once the friction
term $3H\dot{\theta}$ in the equation of motion for the phase $\theta $ becomes
comparable with the 'force' term $m_{\theta }^{2}\theta $, i.e. at the moment
$t^{\ast }$ when
\begin{equation}
H(t^{\ast })\simeq m_{\theta }.  \label{tstar}
\end{equation}%
Taking into account the expression for the Hubble parameter $H=\sqrt{\frac{%
8\pi }{3}\frac{\rho }{M_{P}^{2}}},$ and the connection of the temperature
with time $T=\left( \frac{45}{32\pi ^{2}g_{\ast }}\right) ^{1/4}\sqrt{\frac{%
M_{P}}{t}},$ one obtains the time moment and the temperature of walls
nucleation in the early Universe
\begin{equation}
t^{\ast }=\sqrt{\frac{\pi }{8m_{\theta }^{2}}};\quad T^{\ast }=\left( \frac{%
45}{4\pi ^{3}g_{\ast }}\right) ^{1/4}\sqrt{M_{P}m_{\theta }}.  \label{twall}
\end{equation}%
Applying these formulas to (\ref{lhor}) and taking into account that the
seeking scale can be expressed in the same manner as (\ref{radtemp}), with $%
T^{\ast }$ instead of $T$, one can easily find the size of wall when it
enters horizon
\begin{equation}
R_{2}(N)=\frac{1}{2}\left( \frac{T_{0}L_{U}}{C}\right)
^{2}e^{2(N-N_{U})},\quad C=\left( \frac{45}{4\pi ^{3}g_{\ast }}\right) ^{1/4}%
\sqrt{M_{P}}.  \label{limit2}
\end{equation}%
Thus the maximal comoving size of a spherical wall at FRW epoch seeded at
$e$-folding number $N$, is determined from the condition
\begin{equation}
R(N)=\min (R_{1}(N),R_{2}(N)).  \label{maxcond}
\end{equation}
The condition (\ref{maxcond}) constrains the size of a spherical wall so
that its contribution into the local density does not dominate when it enters
cosmological horizon.

Indeed, for a sufficiently big wall, of radius exceeding $R_{lwd}=\frac{3}{%
8\pi}\frac{M_p^2}{\sigma}$, given by the constraint (\ref{limit1}), its pieces
start to dominate over radiation energy within the cosmological horizon, while
the wall is still causally disconnected as a whole. Therefore the local
expansion rate in a causally connected area dominated by a piece of the wall
turns to be $a_{loc}(t)\propto t^2$. Such superluminal expansion makes the
successive evolution of this region elusive for the distant observer. It puts
the upper limit (\ref{limit1}) on the maximal mass of BHs that can be formed by
the considered mechanism from a spherically symmetric wall. However account of
the non-perfect initial form of the wall and of the effect of crinkle
flattening, considered in the next section, offers the possibility to avoid the
constraint (\ref{limit1}) even for supermassive walls. Then the condition
(\ref{maxcond}) can be relaxed and the maximal size of wall can always be given
by the size it has, when enters the horizon. Local crinkle decay dominance (see
the next section) leads, however, to deviation of local expansion law from the
pure RD case, what modifies the definition (\ref{limit2}) for the maximal size
of wall.

According to the mechanism under consideration, the energy of the wall with
surrounding particles $E_{w}$ will be converted into a BH of mass:
\begin{equation}
M_{BH}(N)\approx E_{w}  \label{BHMass}
\end{equation}%
This happens when the gravitational radius $r_{grav}=2E_{w}/M_{P}$ is larger
than the minimal size of contraction of the collapsing wall. This minimal
size of contraction does not exceed a few times the wall's width.

Using formulas (\ref{rsphere}), (\ref{KN}) and (\ref{BHMass}) one can
calculate the desired distribution of intermediate BHs in the vicinity of
the most massive one.

In Fig. \ref{massdist} following \cite{jetp} we present the mass spectrum of BHs
produced by the described mechanism. Such spectrum strongly depends on the parameters of the
underlying field model. Here these parameters are chosen in such a way that the
average number of BHs with mass $\sim 10^{4}$ $M_{\odot }$, is about $10^{11}$ what
coincides approximately with a number of galaxies in the visible Universe. 
It makes the difference of the model in Fig. \ref%
{massdist} from the results of \cite{jetp}.  

The mass distribution of intermediate BHs around the most massive one is
presented in Fig. \ref{fig:radius} for the same parameters of the potentials
(\ref{V1}) and (\ref{L1}) as have been chosen to produce the distribution Fig.
\ref{massdist}. Actually we always deal with a cluster of BHs centered by the
most massive one. This BHs cluster contains from the beginning $\sim 0.3\times 10^{19}$
BHs of smallest masses about $\sim 10^{-8}M_{\odot }$ and massive BH with mass
$\sim 2.7\times 10^{4}M_{\odot }$ in its center. The total mass of the BHs in the
cluster is $\sim 0.3\times 10^{11}M_{\odot }$ and is comparable with the mass of
a baryon component in a galaxy. Analysis of Fig. \ref{fig:radius}
reveals that average density decreases quickly with size growing. For example,
average density at the scale 1000 parsec in $10^{4}$ times smaller comparing
with average density in the center, at the scale of order 10pc.

\section{\label{crinkles} Possible role of crinkles}

Let us consider the problem of squaring of a wall surface just after the wall
formation. By definition, the spatial position of any wall is determined by the
surface with phase value $\theta =\pi $. Let we have an approximately flat
surface determined as space points containing the phase $\theta =\pi$ of size
$1/H_{e}^{2}$ at the $e$-folding number $N.$ Then, one $e$-folding later, the
surface will be divided into $1/e^{2}$ parts. Approximately half of them
embrace new value of phase $\theta >\pi $ while the remaining half acquires the
values of phase $\theta <\pi$. The total surface of interest, with the phase
$\theta =\pi ,$ appears to be warped and increases by a factor $\varsigma $
varying in the range $\approx 1$--$2.$ The following cosmological expansion
increases the sizes of crinkles exactly like total size of a closed wall. If
this process lasts for, say, $\tilde{N} $ $e$-foldings, the surface increases
by a factor $\varsigma ^{\tilde{N}}$ in addition to the standard inflationary
expansion by a factor $e^{\tilde{N}} $.

Here some explanation is necessary: In the course of inflation the warping goes
to smaller scales, so that smoothing of large crinkles beyond the horizon is
accompanied by creation of small crinkles within it. It should be reminded that
in this process not walls, but physical conditions for their origin are
created. This fractal-like process of crinkles formation terminates when the
size of the wall becomes comparable with the wall width $d=1/m_{\theta }\neq
0$. The crinkles of sizes smaller than the wall width $d$ will be smoothed out
during the period of wall formation and their effect could be ignored in the
analysis of successive wall evolution. Hence our direct aim is to determine the
$e$-folding $N_{d}$ at which crinkles of size $d$ are still produced.

The scale of the crinkles produced at $e$-folding number $N_{d}$ is
approximately equal to the wall width $d$
\begin{equation}
d\approx l_{crinkle}=r_{U}(t^{\ast })e^{N_{d}-N_{U}}=L_{U}\frac{T_{0}}{
T^{\ast }}e^{N_{d}-N_{U}}.  \label{d}
\end{equation}
This formula immediately leads to the number of $e$-foldings we need:
\begin{equation}
N_{d}=N_{U}+\ln \left[ \frac{T^{\ast }d}{T_{0}L_{U}}\right].  \label{Nd}
\end{equation}
Therefore the mass of a wall just after its nucleation is given by the
expression
\begin{equation}
E_{w}\approx \sigma 4\pi R\left( N\right) ^{2}\varsigma ^{N-N_{d}}.
\label{Ewall}
\end{equation}
Choosing the realistic numbers $N-N_{d} =10$--$20$ and $\varsigma =1.5$, one
can see that the effective mass of the wall increases at the moment of its
formation by a factor $\sim 10^2$--$10^3$. Also one has to note that the
size $R_1(N)$ we have calculated in the previous section should be initially
corrected by a factor $\varsigma^{\frac{1}{3}(N_d-N)}$ once the effects of
crinkles are taken into account.

The internal tension in the wall leads to its local flattening with the
growth of the horizon. The released energy is converted into the kinetic
energy of the wall and into an energy of outgoing waves, which are
predominantly the waves of that scalar field that produces the domain walls.
In our case, it is the phase $\theta $ that acquires dynamical sense of the
angular component of field $\varphi$.

A typical momentum $k$ of the waves is of the order of the inverse wall
width $1/d$, hence $k\sim m_{\theta }$ and the waves are necessary
semirelativistic. Their kinetic energy decreases due to the cosmological
expansion. So particles of mass $m_{\theta }$ are concentrated in the
vicinity of the wall. These particles can further contribute to the process
of BH formation. The situation can become more complex if the particles are
unstable. In this case, the products of their decay are relativistic and
most of the time, they escape capture by the gravitational field of the
wall. The main immediate effect of these relativistic products is the local
heating of the medium, surrounding the region of BH formation. Their
concentration in this region and successive derelativization can later
provide their possible contribution to the formation of BHs and BH clusters.

The density of these particles is diluted by cosmological expansion, whereas
the wall kinetic energy is mainly redshifted, making the expression (\ref%
{Ewall}) approximately valid with $\varsigma \sim 1$ in the period of BH
formation, when the whole wall enters the horizon.

However, the contribution of crinkles can be significant, in the period when
the wall enters the horizon, and its successive evolution proceeds so quickly
that the effects of wall flattening are not suppressed by the cosmological
expansion. In this period the wall flattening and crinkle decays transform the
energy, stored in the smaller-scale crinkles, into the energy of scalar field
waves and the kinetic energy of the wall. These products of wall flattening
dominate in the local energy density, thus preventing the superluminal
expansion of the considered region. If the effect of crinkles is sufficiently
strong, the wall dominance may not arise, since the energy released in the
flattening is always dominant over that of the negative pressure of the
flattened wall within the horizon. The RD (or MD) regime of dominance of the
energy, released in the flattening, implies the difference between the
expansion law within the considered region and the general expansion; however,
being subluminal, it continues until the whole wall enters the horizon.

The longer the period of crinkle-decay-products dominance is, the larger is
the overdensity in this region, as compared with the mean cosmological
density, $(\delta \rho /\rho )\gg 1$, when the wall enters the horizon. This
difference provides the separation of this region from the general expansion
and the effective BH formation in it, similar by to the PBH formation on the
RD stage, considered by \cite{Novikov79} (for a review see \cite%
{Khlopov85,klopmop}). This effects can modify the numerical values in BH
distributions presented in the previous section. A special study is,
however, necessary to take it into account.

\section{\label{discussion} Discussion}

It should be obvious that the mechanism proposed in this article, which
explains the formation of PBHs by the pregeneration of objects that build up
the mass of SMBHs in the centre of the galaxy, works in reality by
production of BHs with smaller masses that accompany the most massive BH. An indirect hint~\cite%
{mkr} that a situation of this sort can arise has been found recently in the
centre a rather famous blazar, Mkr 501. Namely the combination of
observations of the host galaxy allowing a direct measurement of the mass of
the central BH in Mkr 501 and those of the modulation of high energy
emission might be brought into agreement by assuming a binary system of
highly unequal SMBHs hosted in the centre of this blazar. Thus, having BHs
of intermediate masses accompanying the most massive ones, which are capable
of coupling into binary systems, turns out not to be unrealistic.

We investigated in detail the primordial spatial distribution of BHs with
masses in the range $\sim 10^{-9}$-$10^5 M_{\odot}$. It was shown that such BHs
could concentrate around a massive one in the scale of up to 10 Mpc. The total
mass of a group of these primordial BHs can be of the order of a mass of
ordinary galaxy clusters. The number of such a groups could be made to coincide
approximately with a number of galaxy clusters. Their formation begins at a
very early stage, which is characterized by the temperature $T^* \sim \sqrt{M_P
m_{\theta}}\sim 10^5$GeV; for the most massive BHs, it may continue even after
the period of Big Bang nucleosynthesis, but stop in any case long before star
formation. Being compatible with the constraints on the average primordial
light element abundance and with the averaged restrictions on the distortions
of the CMB black-body spectrum, the formation of PBH clusters around the most
massive BH may lead to local peculiarities of pregalactic chemical composition,
and to specific effects in the spectrum, angular distribution and polarization
of small-scale CMB fluctuations, accessible to the observational test. After
recombination of hydrogen at the modern matter-dominated stage, baryons are
influenced by the gravitational potential of a primordial BH group, which
facilitates galaxy formation. The proposed model can offer the alternative to
the standard picture of galaxy formation by the slow growth of baryon--density
fluctuations.

To illustrate our approach we considered the simple model of a PNG field in
an inflationary Universe. The physically self-consistent realization of the
proposed scenario should imply more sophisticated models. Such a realization
should provide the compatibility of this scenario with the observational
constraints. Namely, the large-scale fluctuations of the PNG field should be
suppressed so as not to cause too large CMB anisotropy. Another problem of a
realistic scenario will be to satisfy the constraint on the oscillations of
the energy density of the PNG field, playing the role of cold dark matter,
or to provide a sufficiently rapid decay of PNG particles, if the parameters
of the chosen model lead to the contradiction with this constraint.

At the end of the discussion it should be noted that the existence of such
systems of primordial BHs correlated in space offers a new element to the
standard lambda--CDM model. It could add additional power to the spectrum of
density fluctuations on small scales, where the measurements of the Ly$%
\alpha $ forest are relevant \cite{sperg}. It is important that the described
structures of PBHs do not evaporate significantly due to the large amount of
their constituents. This fact makes the clusters of PBHs very favorable to
being gravitationally bounded during the epoch of galaxy formation, up to the
galaxy clusters scales.

\section*{Acknowledgements}

The work of M.Kh and S.R. was performed in the framework of the State Contract
40.022.1.1.1106 and was partially supported by the RFBR grant 02-02-17490 and
grant UR.02.01.008. (Univer. of Russia).

\end{document}